\documentclass[a4paper, 10pt, conference]{ieeeconf}      

\IEEEoverridecommandlockouts                              
\usepackage{epsfig} 

\title{\LARGE \bf
Learning in cognitive systems with autonomous dynamics
}

\author{ \parbox{3 in}{\centering Claudius Gros and Gregor Kaczor\\
         Institute of Theoretical Physics\\
         J.W. Goethe University\\
         60054 Frankfurt/Main, Germany\\
         {\tt\small http://itp.uni-frankfurt.de/$\sim$gros}}
        }

\begin{document}

\maketitle
\thispagestyle{empty}
\pagestyle{empty}

\begin{abstract}

The activity patterns of highly developed cognitive systems like the 
human brain are dominated by autonomous dynamical processes, that
is by a self-sustained activity which would be present even in the 
absence of external sensory stimuli.

During normal operation the continuous influx of external stimuli
could therefore be completely unrelated to the patterns generated 
internally by the autonomous dynamical process. Learning of 
spurious correlations between external stimuli and autonomously
generated internal activity states needs therefore
to be avoided.

We study this problem within the paradigm of transient state
dynamics for the internal activity, that is for an autonomous
activity characterized by a infinite time-series of transiently 
stable attractor states. We propose that external stimuli
will be relevant during the sensitive periods, the transition period
between one transient state and the subsequent semi-stable attractor.
A diffusive learning signal is generated unsupervised whenever
the stimulus influences the internal dynamics qualitatively.

For testing we have presented to the model system stimuli corresponding
to the bar-stripes problem and found it capable to perform
the required independent-component analysis on its own, 
all the time being continuously and autonomously active.
\end{abstract}

\section{INTRODUCTION}

It is well known that the brain has a highly developed
and complex self-generated dynamical neural activity. 
We are therefore confronted with a dichotomy when
attempting to understand the overall functioning of
the brain or when designing an artificial cognitive
system: A highly developed cognitive system, such as
the brain, is influenced by sensory input but is not driven
directly by the input signals. The cognitive system needs 
however this sensory information vitally for 
adapting to a changing environment and survival.

In this context we then want to address a two-fold goal:
\begin{itemize}
\item Can we formulate a meaningful paradigm for the self-sustained
      internal dynamics of an autonomous cognitive system?
\item How is the internal activity process influenced by sensory signals,
      {\it viz} which are the principles for the respective learning processes?
\end{itemize}
With respect to the first problem we note that an increasing
flux of experimental results from neurobiology supports the notion of
quasi-stationary spontaneous neural activity in the cortex
\cite{abeles95,ringach03,kenet03,damoiseaux06,Honey07}.
It is therefore reasonable to investigate possible set-ups
for artificial cognitive systems centrally based on the
notion of internally generated spontaneous transient states,
as we do in the present investigation.

\begin{figure}[tb]
\centerline{
\includegraphics[width=0.45\textwidth]{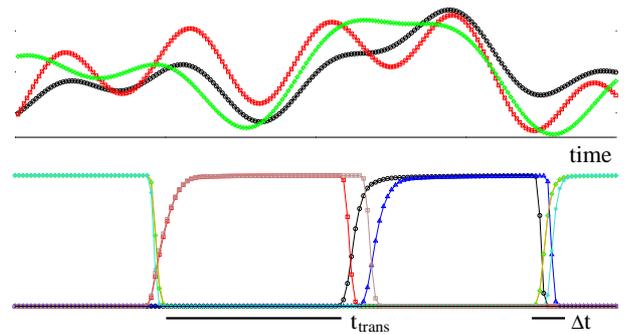}
           }
\caption{Schematic illustration of general or fluctuating
activity patterns (top) and of transient-state
dynamics (bottom), which is characterized by
typical time scales $t_{trans}$ and $\Delta t$
for the length of activity-plateau and of the
transient period respectively.}
\label{figure_trans_activity}
\end{figure}

\section{GENERAL OUTLINE}

We will present the model and the issues being studied
in two steps, starting with a general overview,
discussing technical details in a second step.

\subsection{Transient-state and competitive dynamics}

There are many notions of complex system theory 
characterizing the long-term behavior of a dynamical
system~\cite{grosBook07}, such as regular vs.\ chaotic
behavior. The term `transient-state dynamics'
refers, on the other hand, to the type of activity 
occurring on
intermediate time scales, as illustrated in
Fig.~\ref{figure_trans_activity}. A time
series of semi-stable activity patterns,
also denoted transient attractors, is
characterized by two time scales.
The typical duration $t_{trans}$ of
the activity plateaus and the typical time 
$\Delta t$ needed to perform the transition from
one semi-stable state to the subsequent one.
The transient attractors turns into
stable attractors in the limit $t_{trans}/\Delta t\to\infty$.

Transient state dynamics is intrinsically 
competitive in nature. When the current transient
attractor turns unstable the subsequent transient
state is selected by a competitive process.
Transient-state dynamics is a form of `multi-winners-take-all'
process, with the winning coalition of dynamical
variables suppressing all other competing activities.

Similar processes have been proposed to be relevant
for various neural functionalities.
Edelman and Tononi \cite{edelman00,edelman03} argue that
`critical reentrant events' constitute transient
conscious states in the human brain. These
`states-of-mind' are in their view semi-stable global activity states
of a continuously changing ensemble of neurons, the `dynamic core'.
This activity takes place in what Dehaene and Naccache \cite{dehaene03}
denote the `global workspace'. The global workspace serves, in the
view of Baars and Franklin \cite{baars03},
as an exchange platform for conscious experience 
and working memory. Crick and Koch \cite{crick03} 
and Koch \cite{koch04} have
suggested that the global workspace is made-up of
`essential nodes', i.e. ensembles of neurons responsible
for the explicit representation of particular aspects
of visual scenes or other sensory information.

\begin{figure}[tb]
\centerline{
\includegraphics[width=0.45\textwidth]{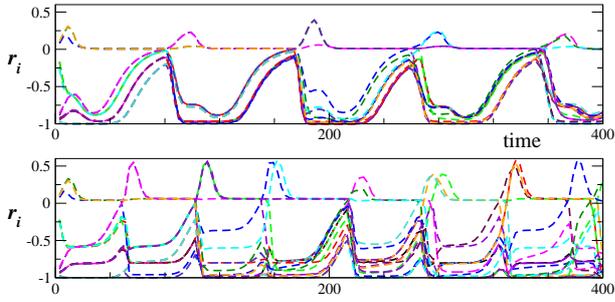}
           }
\caption{The growth rates $r_i(t)$ generating
an internal transient state-dynamics via 
Eq.~(\ref{eq_x_dot}). The two examples differ in the
functional dependence of the $r_i(t)$ on
the $x_j(t)$. The top graph corresponds to
a system having sensitive periods, the bottom
graph to a system without distinctive sensitive
periods.}
\label{figure_sensitive_periods}
\end{figure}

\subsection{Competitive dynamics and sensitive periods}

Within the setting of a generalized neural network
we utilize a continuous-time formulation with
rate-encoding neural activity centers, 
characterized by normalized activity levels
$x_i\in[0,1]$. One can then define, quite generally,
via
\begin{equation}
\dot x_i = 
\left\{
\begin{array}{rl}
(1-x_i)\,r_i & (r_i>0) \\
x_i\, r_i & (r_i<0) 
\end{array}
\right.  
\label{eq_x_dot}
\end{equation}
the respective growth rates $r_i$.
Representative time series of growth rates $r_i$
are illustrated in Fig.~\ref{figure_sensitive_periods}.
When the $r_i>0$ the respective neural 
activity $x_i$ increases, approaching rapidly the
upper bound, as illustrated in
Fig.~\ref{figure_trans_activity}; when
$r_i<0$ it decays to zero. The model is
specified \cite{gros05,grosNJP07},
by providing the functional dependence 
of the growth rates with respect
to the set of activity-states $\{x_j\}$.

During the transition periods many, if not all,
neurons will enter the competition to become
a member of the new winning coalition. The
competition is especially pronounced whenever
most of the growth rates $r_i$ are small in magnitude,
with no small subset of  growth rates dominating the
all the others. Whether this does or does not happen 
depends on the specifics of the model set-up,
in Fig.~\ref{figure_sensitive_periods} 
two cases are illustrated (upper/lower graph
in Fig.~\ref{figure_sensitive_periods}). 
In the first case the competition for the next 
winning coalition is restricted
to a subset of neurons, in the second case the 
competition is network-wide.

When most neurons participate in the competition 
process for a new winning
coalition the model will have `sensitive periods'
during the transition times and it will able
to react to eventual external signals.

\begin{figure}[tb]
\centerline{
\includegraphics[width=0.45\textwidth]{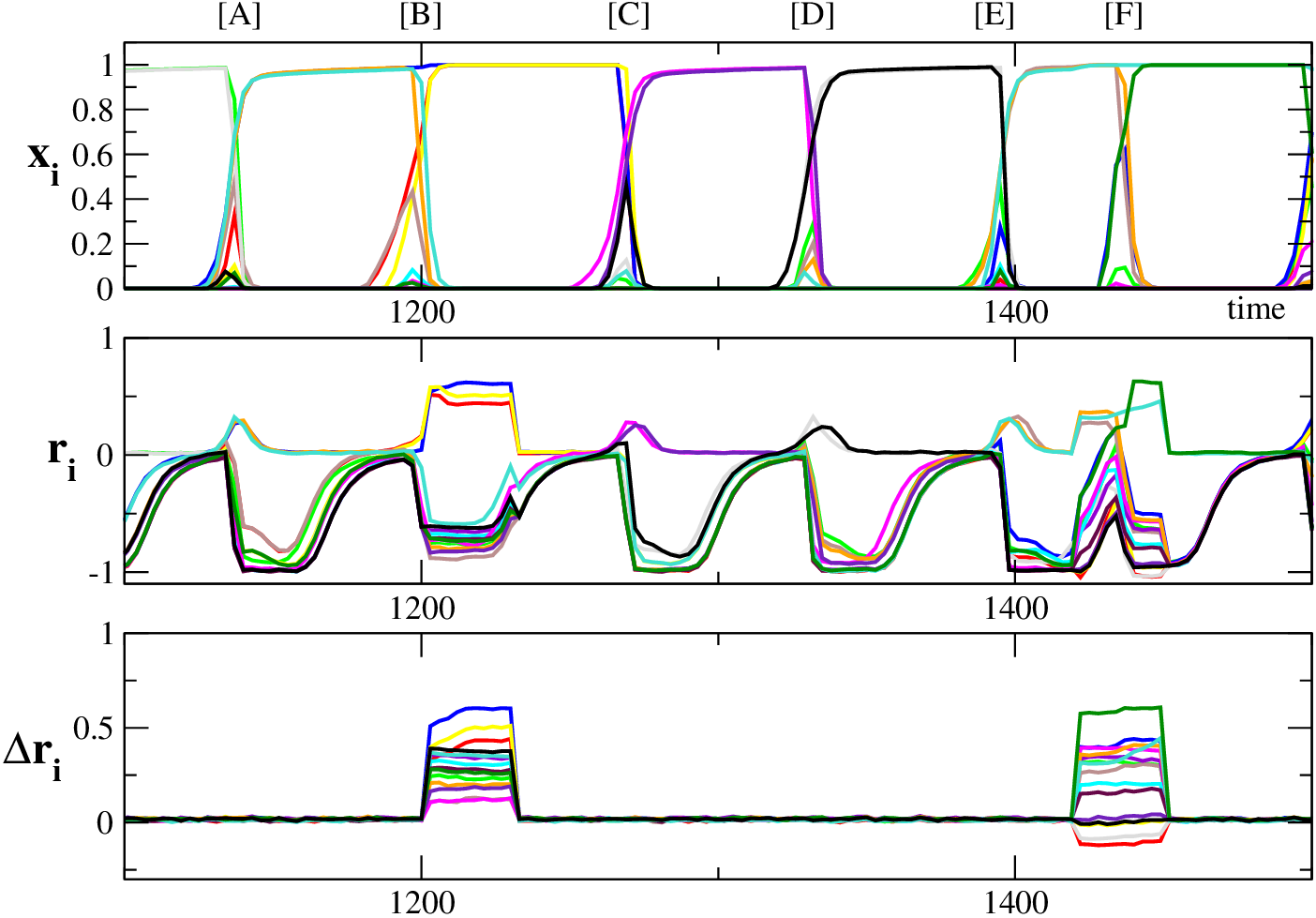}
           }
\caption{Result for the activities $x_i(t)$, the
growth rates $r_i(t)$ and the input signals
$\Delta r_i(t)$ from a simulation of a system
containing $N=15$ sites (color coding). The
time series of winning coalition is (from left
to right):
(1,5) ${[A]\atop\longrightarrow}$ (2,9,12) 
      ${[B]\atop\longrightarrow}$ (0,2,3)
      ${[C]\atop\longrightarrow}$ (8,10)   
      ${[D]\atop\longrightarrow}$ (5,14)
      ${[E]\atop\longrightarrow}$ (4,9,12) 
      ${[F]\atop\longrightarrow}$ (12,13),
where $[..]$ corresponds to the transition-label
given in the graph.}
\label{figure_ARB_15}
\end{figure}

\subsection{Sensitive periods and learning}

So far we have discussed in general terms the properties
of isolated models exhibiting a self-sustained dynamical behavior
in terms of a never-ending time series of semi-stable
transient states, as illustrated
in Fig.~\ref{figure_sensitive_periods},
using rate-encoding equations specified
in previous work \cite{gros05,grosNJP07}.

The importance of sensitive periods
comes when this model is coupled to a stream of
sensory input signals. It is reasonable to assume, 
that external input signals
will contribute to the growth rates $r_i$ 
via
\begin{equation}
r_i \to r_i + \Delta r_i,
\label{eq_delta_r}
\end{equation}
where the $\Delta r_i$ encode the influence of the input
signals. Let us furthermore assume that the
input signals are suitably normalized, such that
\begin{equation}
\Delta r_i \simeq \left\{
\begin{array}{rl}
0.5 & \mbox{(active input)} \\
0 & \mbox{(inactive input)} 
\end{array}
\right.  .
\label{eq_delta_r_magnitude}
\end{equation}
For the transient states
the $r_i\approx -1$ for all sites not
forming part of the winning coalition
and the input signal will therefore not destroy 
the transient state. With the normalization
given by Eq.~(\ref{eq_delta_r_magnitude}) the 
total growth rate $r_i+\Delta r_i$ will remain 
negative for all inactive sites.
The input signal will however
enter into the competition for the next
winning coalition during a sensitive period, 
providing an additional boost for the 
respective neurons. 

This situation is exemplified in
Fig.~\ref{figure_ARB_15}, where we present
simulation-results for a system containing 
$N=15$ neurons subject to two sensory 
inputs $\Delta r_i(t)$. The self-generated
time series of winning coalitions is both
times redirected by the strongest component of
the sensory input, the details depending on whether
the input signal partially overlaps with the
current winning coalition or not.

We then have a model system in which there
are well defined time-windows suitable 
for the learning of correlations between the 
input signal and the intrinsic dynamical activity, 
namely during and shortly after a transition 
period, that is the sensitive period. A possible
concrete implementation for this type of
learning algorithm will be given further below.

This set-up would allow the system also to react to
an occasional very strong input signal having
$\Delta r_i>1$. Such a strong signal
would suppress the current transient state 
altogether and impose itself. This possibility of 
rare strong input signals is evidently important for
animals and would be, presumably, also helpful
for an artificial cognitive system.

\subsection{Diffusive learning signals}

Let us return to the central problem inherent
to all system reacting to input signals having
at the same time a non-trivial intrinsic dynamical 
activity. Namely when should learning occur, i.e.\
when should a distinct activity center become more
sensitive to a specific input pattern and
when should it suppress its sensibility to
a sensory signal.

The above developed framework of competitive
dynamics allows for a straightforward solution
of this central issue: Learning should occur
then and only then when the input signal makes
a qualitative difference, {\it viz} when the
input signal deviates the transient-state
process. For illustration let us assume that
the series of winning coalitions is
$$
(1,5)\ {[a]\atop\longrightarrow}\ (2,9,12) 
     \ {[a]\atop\longrightarrow}\ (0,2,3)~,
$$
where the index $[a]$ indicates that the
transition is driven by the internal
autonomous dynamics and that the series of
winning coalitions take the form
$$
(1,5)\ {[a]\atop\longrightarrow}\ (2,9,12) 
     \ {[s]\atop\longrightarrow}\ (0,4,10)
$$
in the presence of a certain sensory signal $[s]$. 
Note, that a background of weak or noisy sensory input 
could be present in the first case, but learning should 
nevertheless occur only in the second case.

Technically this is achieved by defining a 
suitable diffusive learning signal\footnote{The 
name `diffusive learning signal' 
\cite{grosBook07} stems from the fact, 
that many neuromodulators are
released in the brain in the intercellular 
medium and then diffuse physically to the surrounding
neurons, influencing the behavior of large
neural assemblies.} $S(t)$. It is activated
whenever any of the input signals $\Delta r_i$
changes the sign of the respective growth
rates during the sensitive periods,
\begin{equation}
\dot S \ \to\  \left\{
\begin{array}{rl}
\phantom{-}\Gamma_{diff}^+ & (r_i>0) \mbox{\ and\ } (r_i-\Delta r_i<0)\\
-\Gamma_{diff}^- & \mbox{otherwise} 
\end{array}
\right. 
\label{eq_dot_S}
\end{equation}
{\it viz} when it makes a qualitative
difference. Here $(r_i-\Delta r_i)$ is
the internal contribution to the growth rate,
i.e.\ the value it would have in the absence
of the input signal $\Delta r_i$ and
the $\Gamma_{diff}^{\pm}>0$ are the 
growth and decay rates for the diffusive
learning signal respectively.
The diffusive learning signal $S(t)$ is a 
global signal and a sum $\sum_i$ over all dynamical
variables is therefore implicit on the
right-hand side of Eq.~(\ref{eq_dot_S}).

\subsection{The role of attention}

The general procedure for the learning of
correlation between external signals and
intrinsic dynamical state for a cognitive
system presented here does not rule out
other mechanisms. Here we concentrate on the
learning algorithm which occurs automatically,
one could say sub-consciously. Active
attention focusing, which is well known in
the brain to potentially shut off a sensory 
input pathway or to enhance sensibility
to it, may very well work in parallel, to
the continuously ongoing mechanism investigated
here.

\begin{figure}[t]
\centerline{
\includegraphics*[width=0.45\textwidth]{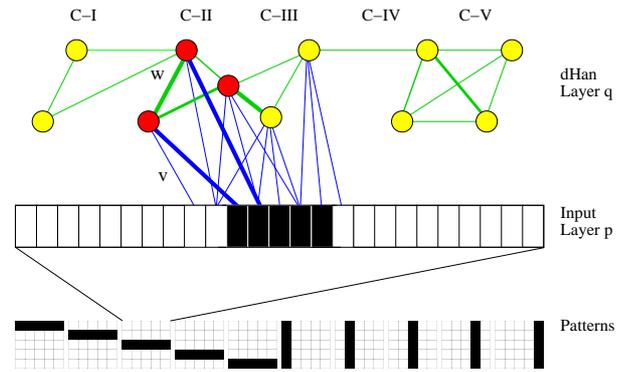}
           }
\caption{
Schematic representation of the information flow from a raw pattern 
(bottom) via the input layer (middle) to the dHan layer (top). \newline
The synaptic strength $v_{ij}^{pq}$ connecting the input with the
dHan layer (denote `v' in the graph) are adapted during the
learning process (illustrated selectively by respective 
thick/thin blue lines in the graph). \newline
The dHan layer consists of active and inactive neurons (red/yellow circles)
connected by intra-layer synaptic weights (denote `w' in the graph).
The topology shows five cliques (denoted C-I to C-V in the graph)
of which C-II is active, as emphasized by the red-color neurons.
        }
\label{fig_twoLayers}
\end{figure}

\section{ASSOCIATIVE THOUGHT PROCESSES}

So far we have described, in general terms, the system
we are investigating, having sensitive periods during
the transition periods of the continuously ongoing
transient-state process, with learning of input
signals regulated by a diffusive learning signal.

Our system consists of two components, as illustrated
in Fig.~\ref{fig_twoLayers}. For the component generating
an infinite time-series of transient state we
employ a dense homogeneous associative network (dHAN).
The dHAN-model has been studied previously 
and shown to generated a time series
of transient states characterized by high
associative overlaps between subsequent winning
coalitions~\cite{gros05,grosNJP07}. The time series might
therefore be interpreted as
an associative thought process, it carries with it
a dynamical attention field \cite{gros05}.

\subsection{Input data-stream analysis}

The input signal acts via Eq.~(\ref{eq_delta_r}) on
the dHAN layer, with the contribution $\Delta r_i$
to the growth rate of the dHAN neuron $i$
given by
\begin{equation}
\Delta r_i^p = \sum_{j} v_{ij}^{pq} x_j^q,
\qquad\quad
\Delta s_i^p = \sum_{j} v_{ij}^{pq} (1-x_j^q),
\label{eq_Delta_r_v_pq}
\end{equation}
where we have denoted now with the superscripts $p/q$
the dHAN- and input-layer respectively. For subsequent
use we also define in Eq.~(\ref{eq_Delta_r_v_pq}) an
auxiliary variable $\Delta s_i^p$ which quantifies
the influence of inactive input-neurons. The task
is now to find a suitable learning algorithm which
extracts the relevant information from the input-data
stream by mapping distinct input-patterns onto
selected winning coalitions of the dHAN layer.
This is the setup typical for an independent
component analysis \cite{hyvarinen00}.

The multi-winners-take-all dynamics in the dHAN
module implies individual neural activities to
be close to 0/1 during the transient states 
and we can therefore define three types of
inter-layer links $v_{ij}^{pq}$ (see
Fig.~\ref{fig_twoLayers}):

\begin{itemize}
\item \underline{active} ({\it `act'})\\
      Links connecting active input neurons
      with the winning coalition of the
      dHAN module. 
\item \underline{orthogonal} ({\it `orth'})\\
      Links connecting inactive input neurons
      with the winning coalition of the
      dHAN module.
\item \underline{inactive} ({\it `ina'})\\
      Links connecting active input neurons
      with inactive neurons of the
      dHAN module.
\end{itemize}
The orthogonal links
take their name from the circumstance that
the receptive fields of the winning coalition
of the target layer need to orthogonalize to
all input-patters differing from the present one.
Note that it is not the receptive field of individual
dHAN-neurons which is relevant, but rather 
the cumulative receptive field of a given winning
coalition.

We can then formulate three simple rules for the 
respective link-plasticity. Whenever
the new winning coalition in the dHAN-layer
is activated by the input layer, {\it viz}
whenever there is a substantial 
diffusive learning signal, 
i.e.\ when $S_{diff}^p$ exceeds a certain
threshold $S_{diff}^c$, the following
optimization procedures should take
place:
\begin{itemize}
\item \underline{active links}\\
      The sum over active links should take
      a large but finite value $r_v^{act}$,
$$
\sum_{x_j^q\,{\rm active}} v_{ij}^{pq}
\bigg|_{x_i^p\,{\rm active}}
 \to\ r_v^{act}~.
$$
\item \underline{orthogonal links}\\
      The sum over orthogonal links should take
      a small value $s_v^{orth}$,
$$
\sum_{x_j^q\,{\rm inactive}} v_{ij}^{pq}
\bigg|_{x_i^p\,{\rm active}}
 \to\ s_v^{orth}~.
$$
\item \underline{inactive links}\\
      The sum over inactive links should take
      a small but non-vanishing value $r_v^{ina}$,
$$
\sum_{x_j^q\,{\rm active}} v_{ij}^{pq}
\bigg|_{x_i^p\,{\rm inactive}}
 \to\ r_v^{ina}~.
$$
\end{itemize}
The $r_v^{act}$, $r_v^{ina}$ and
$s_v^{orth}$ are the target values for the
respective optimization processes.
In order to implement these three rules 
we define three corresponding contributions
to the link plasticities:
\begin{equation}
\begin{array}{rcl}
c_i^{act} & =& \Gamma_v^{act}\,\Theta(x_i^p-x_v^{act})
               \,{\rm Sign}(r_v^{act}-\Delta r_i^p) \\
c_i^{orth} & =& \Gamma_v^{orth}\,\Theta(x_i^p-x_v^{act})
               \,{\rm Sign}(s_v^{orth}-\Delta s_i^p) \\
c_i^{ina} & =& \Gamma_v^{ina}\,\Theta(x_v^{ina}-x_i^p)
               \,{\rm Sign}(r_v^{ina}-\Delta r_i^p) 
\end{array}
\label{eq_back_contributions}
\end{equation}
where the inputs $\Delta r_i^p$ and
$\Delta s_i^p$ to the target layer
are defined by Eq.~(\ref{eq_Delta_r_v_pq}). For the
sign-function ${\rm Sign}(x)=\pm1$ is valid 
for $x>0$ and $x<0$ respectively, 
$\theta(x)$ denotes the Heaviside-step
function in Eq.~(\ref{eq_back_contributions}).
The inter-layer links $v_{ij}^{pq}$ cease
to be modified whenever the total input
is optimal, {\it viz} when no more `mistakes` 
are made \cite{chialvo99}.

\begin{figure}[tb]
\centerline{
\includegraphics[width=0.45\textwidth]{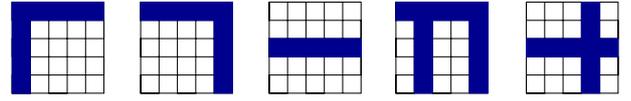}
           }
\caption{Examples of typical input patterns for a
$5\times 5$ bars problem with a probability $p=0.1$
for the occurrence of the individual horizontal or
vertical bars. The problem is non-linear since the
pattern intensity is not enhanced when an elementary
horizontal and vertical bar overlap.}
\label{figure_bars_pattern}
\end{figure}

\begin{table}[b]
\caption{A possible set of parameters for
the inter-layer links, with $\Gamma_{diff}^+=4.0$ and
$\Gamma_{diff}^-=0.08$.
        }
\begin{center}
\begin{tabular}{c|c|c|c}
\hline \hline
$ \Gamma_v^{act} $ \ $ \Gamma_v^{orth} $ \ 
$ \Gamma_v^{ina} $ & $ r_v^{act}       $ \ 
$ s_v^{orth}     $ \ $ r_v^{ina}       $ & 
$ x_v^{act}      $ \ $ x_v^{ina}       $ &
$ S_{diff}^c     $ 
\\
\hline
0.008 \ 0.001 \ 0.001 & 0.9 \ \ \ 0.3 \ \ \ 0.2 & 0.4 \ \ \ 0.2 & 0.25 \\
\hline \hline
\end{tabular}
\end{center}
\label{tab_par_back}
\end{table}

Using these definitions, the link plasticity
may be written as
\begin{eqnarray} 
\label{eq_vdot}
\dot v_{ij}^{pq} & =& \Theta(S_{diff}^p-S_{diff}^c)\,
\Big[\, c_i^{act}\,x_j^q \\
 & +&  c_i^{orth}\,\left(1-x_j^q\right) 
 + c_i^{ina}\,x_j^q\,\Big]~, 
\nonumber 
\end{eqnarray}
where $S_{diff}^c$ is an appropriate threshold for
the diffusive learning signal
$r_v^{act}$ and $r_v^{ina}$ are the desired contributions
to the growth rate for active/inactive postsynaptic
neurons, $\Gamma_v^{act}$ and $\Gamma_v^{ina}$
the respective rates and $x_v^{act}$ and $x_v^{ina}$ the 
critical activity reservoirs defining an 
active/inactive postsynaptic center respectively. 
A suitable set of parameters, which has been used
throughout this work, is given in Table~\ref{tab_par_back}.

%
\begin{figure*}[tb]
\centerline{
\includegraphics[width=0.80\textwidth]{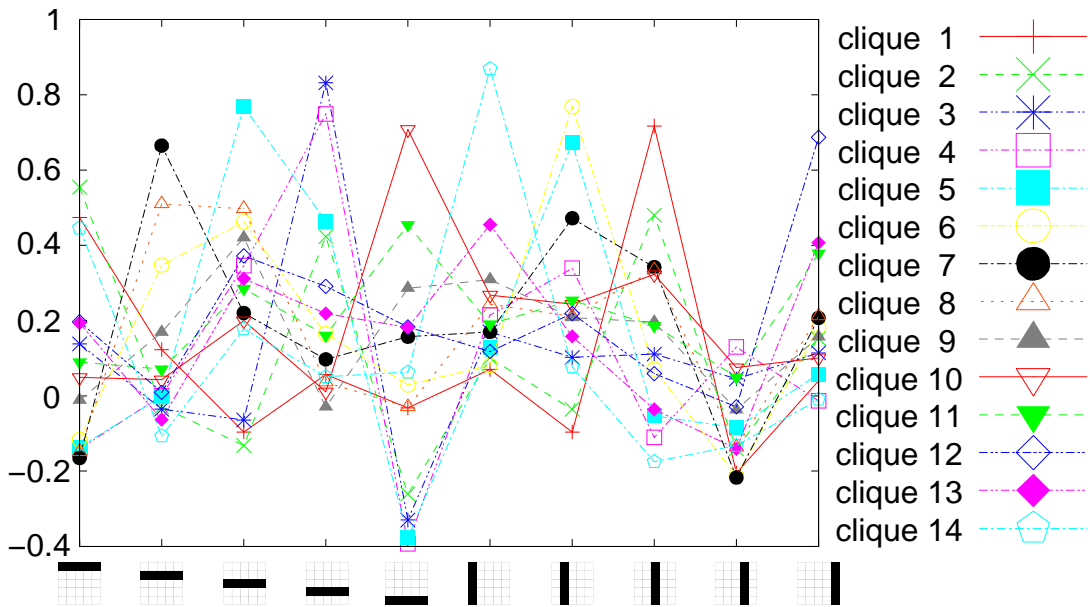}
           }
\caption{The response, as defined by Eq.~(\ref{eq_clique_rec_patt}),
for the 14 winning coalitions in the dHAN layer
(compare Fig.~\ref{figure_cRF})
to the ten reference patterns, {\it viz} the 5 horizontal bars 
and the 5 vertical bars of the $5\times5$ input field.}
\label{figure_cRP_graph}
\end{figure*}

We note, that a given interlayer-link
$v_{ij}^{pq}$ is in general subject to
competitive optimization from the three
processes (act/orth/ina). Averaging
would occur if the respective learning
rates $\Gamma_v^{act}$/$\Gamma_v^{orth}$/$\Gamma_v^{ina}$
would be of the same order of magnitude.
It is therefore necessary, that
$$
\Gamma_v^{act}\ \gg\ \Gamma_v^{orth},
\qquad\quad
\Gamma_v^{act}\ \gg\ \Gamma_v^{ina}~.
$$

\subsection{Homeostatic normalization}

It is desirable that the interlayer connections 
$v_{ij}^{pq}$ neither grow unbounded with
time (runaway-effect) nor disappear into 
irrelevance. Suitable normalization procedures
are therefore normally included explicitly into the 
respective neural learning rules and are present implicitly
in Eqs.~(\ref{eq_back_contributions}) and
Eq.~(\ref{eq_vdot}). 

The strength of the input-signal is optimized
by Eq.~(\ref{eq_vdot}) both for active as well as
for inactive $p$-layer neurons, a property referred to as
fan-in normalization. Eqs.~(\ref{eq_back_contributions}) 
and (\ref{eq_vdot}) also regulate the overall
strength of inter-layer links emanating from
a given $q$-layer neuron, a property called
fan-out normalization. 

Next we note, that the timescales for the intrinsic autonomous
dynamics in the dHAN-layer and for the input signal could in
principle differ substantially. Potential
interference problems can be avoided when
learning is switched/one very fast, {\it viz} when
activation and when the decay rate
$\Gamma_{diff}^\pm$ are larger, i.e.\
when $1/\Gamma_{diff}^\pm$ is smaller
than both the typical time scales of the
input and of the self-sustained dynamics.

\section{THE BARS PROBLEM}

A cognitive system needs to autonomously extract
meaningful information about its environment from its
sensory input data stream via signal separation
and features extraction. The identification of recurrently
appearing patterns, i.e.\ of objects, in the background of fluctuation
combinations of distinct and noisy patterns constitutes a
core demand in this context. This is the domain of
the independent component analysis \cite{hyvarinen00} 
and blind source separation \cite{blindSource},
which seeks to find distinct representations of statistically
independent input patterns.
 
In order to test our system made-up by an input-layer coupled
to a dHAN layer, as illustrated in Fig.~\ref{fig_twoLayers},
we have selected the bars problem \cite{barsProblem}.
The bars problem constitutes a standard non-linear reference 
task for the feature extraction via an independent component 
analysis for a $L\times L$ input layer. Basic patterns 
are the $L$ vertical and $L$ horizontal bars. The 
individual input patterns are made-up of a non-linear superposition
of the $2L$ basic bars, containing with probability $p=0.1$ 
any one of them, as shown in Fig.~\ref{figure_bars_pattern}.

\subsection{Simulation results}

For the simulations we presented to the system about
$N_{patt}\approx200$ randomly generated $5\times 5$ input 
patterns of the type shown in Fig.~\ref{figure_bars_pattern},
including a small noise-level of about $5\%$. The individual
patterns lasted $T_{patt}=30$ with about $T_{inter}=190$
for the time in between two successive input signals. These
time-scales are to be compared with the time-scale of
the autonomous dHAN-dynamics illustrated in
the Figs.~\ref{figure_sensitive_periods} and
\ref{figure_ARB_15}, for which the typical stability-period
for a transient state is about $t_{trans}\approx 70$.
This implies that we presented the system sparse (in time) 
sensory input. We also note that the number of training patterns 
in our simulation is exceedingly small,
standard neural algorithms use routinely $\sim 10^4$
and more training patterns \cite{triesch07}.

The results for the simulations are presented in
Figs.~\ref{figure_cRP_graph} and \ref{figure_cRF}.
For the geometry of the dHAN network we used a
simple 15-site chain containing 14
potential winning coalitions, as
illustrated in Fig.~\ref{figure_cRF}, namely
$(0,1)$, $(1,2)$, \dots, $(14,15)$.

In Fig.~\ref{figure_cRP_graph} we present the  response
\begin{equation}
{1\over S(C_\alpha)} 
\sum_{i\in C_\alpha,j} v_{ij}^{pq} x_j^{q,\beta},
\qquad 
\begin{array}{rcl}
&&\alpha=1,..,14 \\
&&\beta = 1,..,10
\end{array}
\label{eq_clique_rec_patt}
\end{equation}
of the 14 potential winning coalitions $C_\alpha$ 
to the 10 basic input patterns
$\{x_j^{q,\beta},j=1,..,25\}$, the isolated bars. Here
$C_\alpha$ denotes the set of sites of the winning-coalition
$\alpha$ and $S(C_\alpha)$ its size, here $S(C_\alpha)=2$.

The individual potential winning coalitions have acquired
in the course of the simulation, via the
learning rule Eq.~(\ref{eq_vdot}),
distinct susceptibilities to the 10 bars
compare Fig.~\ref{figure_cRP_graph}. We note
that the learning is unsupervised and quite fast.
This example-problem is overcomplete, 
there are more potential winning coalitions than 
statistically independent basic pattern. Perfect signal
separation can therefore not be expected. E.g.\ the second-last
vertical bar in Fig.~\ref{figure_cRF} is not resolved.

As the dHAN dynamics is
competitive in nature already relatively small 
differences in the response, Eq.~(\ref{eq_clique_rec_patt}),
may determine the outcome of the competition between
two competing potential winning coalitions. This
circumstance is born-out by the results presented in
Fig.~\ref{figure_cRP_graph}. 

\subsection{Receptive fields}

\begin{figure}[tb]
\centerline{
\includegraphics[width=0.45\textwidth]{receptiveField.epsi}
           }
\bigskip
\centerline{
\includegraphics[width=0.45\textwidth]{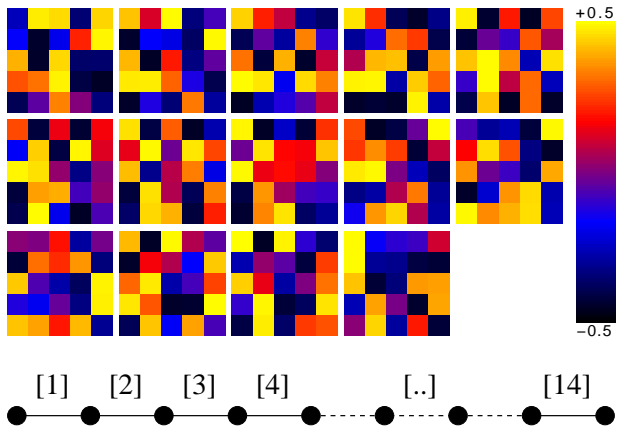}
           }
\caption{Top: The receptive fields, Eq.~(\ref{eq_clique_recp_fields}),
for the 14 winning coalitions $[1],..,[5]$ (first row) 
$[6],..,[10]$ (second row) and $[11],..,[14]$ (third row)
as a function of the $5\times5$ input field, compare
Fig.~\ref{figure_bars_pattern}. For the response of the
14 winning coalitions with respect to the ten reference input patterns 
compare Fig.~\ref{figure_cRP_graph}. Note that the receptive
fields can be both positive as well as negative, see the
color-coding.\newline
Bottom: The geometry of the dHAN layer as
a linear chain. The winning coalitions 
$[i]$ ($i=1,..,14$) are numerated and 
correspond here to two connected 
nearest-neighbor in a linear-chain layout.
        }
\label{figure_cRF}
\end{figure}

The receptive fields
\begin{equation}
{1\over S(C_\alpha)} \sum_{i\in C_\alpha} v_{ij}^{pq},
\qquad \alpha=1,..,14,
\label{eq_clique_recp_fields}
\end{equation}
of the 14 potential winning coalitions are given in
Fig.~\ref{figure_cRF}. The inter-layer synaptic weights
$v_{ij}^{pq}$ can be both positive and negative and
the orthogonalization procedure, Eq.~(\ref{eq_back_contributions}),
results in complex receptive fields. 

Physically separated target neurons, {\it viz} a 
`single-winner-takes-all' set-up, are normally
used for standard analysis neural algorithms
performing an independent component analysis
\cite{hyvarinen00}. The winning coalitions of
the dHAN-layer are however overlapping and
every link $v_{ij}^{pq}$ targets in general more than one
potential winning coalition in the dHAN layer,
in our case two for $i=2,..,14$ (and one for 
$i=1,15$). 

The unsupervised learning procedure,
Eq.~(\ref{eq_vdot}), 
involves therefore a competition
between the active contribution $c_i^{act}$
and the $c_î^{orth}$ and $c_î^{ina}$, as given
by Eq.~(\ref{eq_back_contributions}). The
inter-layer synaptic weights are therefore
only changed to the point necessary for
recognition, but not up to the saturation point,
as evident from the data presented in
Fig.~\ref{figure_cRF}. 
This behavior is consistent with the 
`learning by mistakes' paradigm \cite{chialvo99},
which states that a cognitive system needs to
learn in general only when committing a mistake.

\section{CONCLUSIONS AND FUTURE WORKS}

A cognitive system has its own internal dynamics and 
we studied here the interplay of these self-generated
activity states, the time-series of winning coalitions, 
with the sensory input for the purpose of unsupervised 
feature extraction. We proposed learning to be autonomously 
activated during the transition from one winning 
coalition to the subsequent one.

This general principle may be implemented algorithmically
in various fashion. Here we used a generalized neural
net (dHAN - dense homogeneous associative net) for
the autonomous generation of a time series of associatively
connected winning coalitions and controlled the 
unsupervised extraction of input-features by an autonomously
generated diffusive learning signal.

We tested the algorithm for the bars problem and
found good and fast learning for the case of sparse
temporal input. Preliminary results indicate that the
learning algorithm retains functionality under a wide
range of conditions. We plan to extend the simulations 
to various forms of temporal inputs, especially to 
quasi-continuous input and to natural scene analysis
and to study the embedding of the here proposed concept
within the framework of a full-fledged and
autonomously active cognitive system.



\addtolength{\textheight}{-3cm}   


\begin{thebibliography}{99}

\bibitem{abeles95} M. Abeles {\it et al.},
Cortical activity flips among quasi-stationary states,
{\it PNAS}, vol.  92, 1995, pp 8616-8620.

\bibitem{ringach03} D.L. Ringach, States of mind,
{\it Nature}, vol. 425, 2003, pp 912-913.

\bibitem{kenet03} T. Kenet, D. Bibitchkov, M. Tsodyks, A. Grinvald
                  and A. Arieli,
Spontaneously emerging cortical representations of visual attributes,
{\it Nature}, vol. 425, 2003, pp 954-956.

\bibitem{damoiseaux06}
J.S. Damoiseaux, S.A.R.B. Rombouts, F. Barkhof, P. Scheltens, 
C.J. Stam, S.M. Smith and C.F. Beckmann,
Consistent resting-state networks across healthy subjects,
{\it PNAS}, vol. 103, 2006, pp 13848-13853.

\bibitem{Honey07}
C.J. Honey, R. K\"otter, M. Breakspear and Olaf Sporns,
Network structure of cerebral cortex shapes functional 
       connectivity on multiple time scales,
{\it PNAS}, vol. 104, 2007, pp 10240-10245.

\bibitem{grosBook07} C. Gros,
{\it Complex and Adaptive Dynamical Systems, A Primer},
Springer; 2007 (in press).

\bibitem{edelman00} G.M. Edelman and G.A. Tononi,
{\it A Universe of Consciousness},
New York: Basic Books; 2000.

\bibitem{edelman03} G.M. Edelman,
Naturalizing consciousness: A theoretical framework,
PNAS, vol. 100, 2003, pp 5520-5524.

\bibitem{dehaene03} S. Dehaene and L. Naccache,
Towards a cognitive neuroscience of consciousness:
basic evidence and a workspace framework,
{\it Cognition}, vol. 79, 2003, pp 1-37.

\bibitem{baars03} B.J. Baars and S. Franklin,
How conscious experience and working memory interact,
{\it Trends Cog. Science}, vol. 7, 2003, pp 166-172.

\bibitem{crick03} F.C. Crick and C. Koch,
A framework for consciousness,
{\it Nature Neurosci.}, vol. 6, 2003, pp 119-126.

\bibitem{koch04} C. Koch,
{\it The Quest for Consciousness - A Neurobiological Approach},
Robert and Company; 2004.

\bibitem{gros05} C. Gros,
Self-Sustained Thought Processes in a Dense Associative Network,
{\it in KI 2005}, U. Furbach (Ed.),
{\it Springer Lecture Notes in Artificial Intelligence}, vol. 3698,
2005, 366-379; also available as
http://arxiv.org/abs/q-bio.NC/0508032.

\bibitem{grosNJP07} C. Gros,
Neural networks with transient state dynamics,
{\it New J. Phys.}, vol. 9, 2007, 109.

\bibitem{hyvarinen00} A. Hyv\"arinen, and E. Oja,
Independent component analysis: Algorithms and applications,
{\it Neural Networks}, vol. 13, 2000, 411-430.

\bibitem{chialvo99} D.R. Chialvo and P. Bak,
Learning from mistakes.
{\it Neuroscience}, vol. 90, 1999, 1137-1148.	

\bibitem{blindSource} S. Choi, A. Cichocki, H.M. Park, and S.Y. Lee,
Blind Source Separation and Independent Component Analysis: A Review,
{\it Neural Information Processing}, vol. 6, 2005, 1-57.

\bibitem{barsProblem} P. F\"oldi\'ak,
Forming sparse representations by local anti-Hebbian learning,
{\it Biological Cybernetics}, vol. 64, 1990, 165-170.

\bibitem{triesch07} N. Butko, and J. Triesch,
Learning Sensory Representations with Intrinsic Plasticity,
{\it Neurocomputing}, vol. 70, 2007, 1130-1138.
  





\end{thebibliography}
\end{document}